\documentclass[aps,pre, twocolumn, groupedaddress]{revtex4-1}

\usepackage{graphicx}
\usepackage{dcolumn}
\usepackage{amsmath}    
\usepackage{amssymb}
\usepackage{bm} 
\usepackage{hyperref}
\usepackage{latexsym}
\usepackage{verbatim}
\usepackage{color}
\def\beq{\begin{equation}}
\def\eeq{\end{equation}}

\def\beq{\begin{equation}}                           
\def\eeq{\end{equation}}                           
\def\bea{\begin{eqnarray}}                           
\def\eea{\end{eqnarray}}        

\begin{document}



\title{\textcolor{black}{\Large  Tighter Einstein-Podolsky-Rosen steering inequality based on the sum uncertainty relation}}


\author{Ananda G. Maity$^{1,*}$, Shounak Datta$^{1\dagger}$, A. S. Majumdar$^{1,\ddagger}$}
\email[]{$*$anandamaity289@gmail.com \\ $\dagger$shounak.datta@bose.res.in \\ $\ddagger$ archan@bose.res.in}
\affiliation{$^1$S. N. Bose National Centre for Basic Sciences, JD Block, Sector III, Salt Lake City, Kolkata 700106}

\date{\today}

\begin{abstract}

{\textcolor{black}{We consider the uncertainty bound on the sum of variances of two incompatible
observables in order to derive a corresponding steering inequality. Our steering criterion when
applied to discrete variables yields the optimum steering range for two qubit Werner states in
the two measurement and two outcome scenario. We further employ the derived steering relation for
several classes of continuous variable systems. We show that non-Gaussian entangled states such as the
photon subtracted squeezed vacuum state and the two-dimensional harmonic oscillator state furnish
greater violation of the sum steering relation compared to the Reid criterion as well as the
entropic steering criterion.  The sum steering inequality provides a tighter steering condition
to reveal the steerability of continuous variable states.}}

\end{abstract}


\maketitle

\section{Introduction \label{introduction}}

The idea of nonlocal correlations in quantum mechanics (QM) emanated from the work of  Einstein, Podolsky and Rosen~\cite{EPR} in 1935. Considering a position-momentum-correlated state EPR conjectured that nonlocality is an artefact of the incompleteness of the quantum mechanics. Soon afterwards Schr$\ddot{o}$dinger \cite{Schr1,Schr2} coined the word \textit{entanglement} to describe spatially separated but correlated particles. He also introduced the term \textit{steering}, to describe how the choice of a measurement basis on one side could affect the state on the other. He however,
believed that steering would never be observed experimentally, since there existed on the other side
a definite state even if it is unknown to the local observer (local hidden state(LHS)). 
Much later, Bell~\cite{Bell,chsh} proposed  how correlations in outcomes of joint measurements of observables corresponding to two spatially separated particles sharing an entangled state could be quantified by taking into consideration the requirements of locality and realism. QM violates Bell's inequality, as subsequently verified through several experiments~\cite{aspect}.  Demonstration of 
EPR-steering, on the other hand, was first proposed by Reid~\cite{Reid1}, with subsequent experimental realization by
Ou {\it et al}~\cite{ou} and others~\cite{ss}.

A few years ago, Wiseman {\it et al}~\cite{Wiseman} formulated a unified information
theoretic description of quantum
correlations manifested through entanglement, EPR-steering and Bell-nonlocality  in terms of  information theoretic tasks, and showed a strict hierarchy between these three types of correlation,
{\it viz.},   EPR-steering~\cite{R} lies between Bell nonlocality and entanglement, with the latter being the
weakest. This paved the way for steering inequalities analogous to bell-inequalities  to be formulated to rule out the existence of  LHS models and demonstrate steerability. Experimental manifestation of EPR-steering
using discrete variable Werner states~\cite{Werner} was first performed by Saunders {\it et al}~\cite{Saunders}. Several steering inequalities
have since been  proposed, with the motivation of obtaining stronger or optimal steering criteria corresponding to particular contexts of the number of parties and measurement 
settings~\cite{caval,entropy1,TP}. A necessary and sufficient condition for steering has recently
been obtained by Cavalcanti {\it et al}~\cite{analog}, in the case of bipartite systems with two measurement settings and two 
outcomes per party. 

The quantum uncertainty principle plays a central role in the manifestation of EPR-steering. In the realm of continuous variables, the demonstration of steering by the Reid
formalism~\cite{Reid1}  is based upon the
calculation of inferred variances of quadrature amplitudes. Due to correlations in the observables
of the two parties sharing an entangled state, the product of inferred variances may fall below
the limit obtained through the application of the Heisenberg uncertainty relation (HUR)~\cite{HUR}, thus revealing
steering. It was subsequently realized that the Reid criterion based on the HUR is incapable of demonstrating steerability of several continuous variable states,
notably certain highly entangled and even bell-nonlocal non-Gaussian states. As  entropic
uncertainty relations (EUR)~\cite{entropy,eur} are tighter compared to  the HUR, steering inequalities
based on EURs have since been proposed~\cite{Walborn}. Entropic steering relations provide
stronger conditions for EPR-steering and hence fare better
in revealing steering by non-Gaussian entangled states~\cite{PC-TP}. A more optimal uncertainty bound
is provided by the fine-grained uncertainty relation~\cite{Oppenheim} which has been used to
obtain an even tighter fine-grained steering criterion for continuous variables~\cite{PC}.

Recently, variance based sum uncertainty relations~\cite{Pati,Mac} have received a lot of
attention. Sum uncertainty relations guarantee the lower bound of uncertainty to be non-trivial
for incompatible observables whenever the variance of at least one of the observables is
non-zero, a feature that is lacking in uncertainty relations 
based on the product of variances, such as 
the HUR or the Robertson-Schrodinger uncertainty relation~\cite{RS}. Sum uncertainty relations
therefore, in general,  provide a tighter bound of uncertainty compared to product uncertainty
relations~\cite{tighter}, as has also been realized experimentally~\cite{sumexpt}. Extensions
of the sum uncertainty relations have further been formulated for systems involving multiple 
observables~\cite{summulti}. In the present context, it is hence imperative to enquire what
advantage, if any, would the sum uncertainty relation provide to the corresponding steering 
criterion, compared to say, an HUR based steering inequality. 

With the above motivation, 
in this work we derive a new steering criteria using an uncertainty relation based on the sum of variances~\cite{Pati}. We next apply our steering criteria first to the case of a bipartite system
with two measurements settings and two outcomes. We show that our steering inequality applied
to the Werner state matches the recently derived necessary and sufficient 
condition for steering~\cite{analog} for this setting.  We then move on to study continuous
variable systems where tighter steering conditions based on EURs have been able to reveal steering
by several  non-Gaussian entangled states~\cite{Walborn,PC-TP} in addition to the Gaussian states which
mostly admit steering using the standard HUR. Non-gaussian states generally have a higher degree of
entanglement compared to Gaussian states, and hence, 
 have applications in tests of Bell inequalities, quantum
teleportation, and other quantum information protocols~\cite{nongauss}. We consider various
classes of non-Gaussian states for application of our derived steering inequality. We show that
the sum uncertainty based steering criterion improves upon the steering criteria based
on HUR and EUR for such states

The plan of this paper is as follows. In the next section  we derive a steering inequality from  an uncertainty 
relation based on the sum of variances. In section III we show that our steering condition
matches the necessary and sufficient condition for steering in the case of two-qubit Werner
states. In section IV we apply our steering inequality on 
different classes of continuous variable entangled states such as the two mode squeezed vacuum state 
(TMSV), the photon subtracted TMSV state, and the two-dimensional harmonic oscillator state.
We compare the results obtained with those using the Reid steering criterion and the entropic steering
criterion. Section V is reserved for some concluding remarks.



\section{Steering inequality using sum-uncertainty relation}


EPR-steering was first demonstrated using the HUR by showing that the product of variances of inferred values of the correlated observables is less than the lower bound of uncertainty~\cite{Reid1}. Defining the quadrature phase amplitudes  as
\begin{equation}
\widehat{X}_{\theta} = \frac{\widehat{a}e^{-i\theta}+ \widehat{a}^{\dagger}e^{i\theta}}{\sqrt{2}},\widehat{Y}_{\phi} = \frac{\widehat{b}e^{-i\phi}+ \widehat{b}^{\dagger}e^{i\phi}}{\sqrt{2}}
\end{equation} 
where the operators
$\widehat{a} = \frac{\widehat{X}+i\widehat{P}_{x}}{\sqrt{2}},\widehat{a}^{\dagger} = \frac{\widehat{X}-i\widehat{P}_{x}}{\sqrt{2}} ,
\widehat{b} = \frac{\widehat{Y}+i\widehat{P}_{y}}{\sqrt{2}},\widehat{b}^{\dagger} = \frac{\widehat{Y}-i\widehat{P}_{y}}{\sqrt{2}}$
 obey the bosonic commutation relations. In the presence of correlations
$C_{\theta,\phi} = \frac{\langle\widehat{X}_{\theta}\widehat{Y}_{\phi} \rangle}{\sqrt{\langle\widehat{X}_{\theta}^{2}\rangle\langle\widehat{Y}_{\phi}^{2}\rangle}}$, 
the quadrature amplitude $\widehat{X}_{\theta}$ 
could be inferred by measuring the corresponding amplitude $\widehat{Y}_{\phi}$. Hence, using the HUR,
$\Delta \widehat{X}_{\theta_{1}}^{2}\Delta\widehat{X}_{\theta_{2}}^{2}\geq 1/4$
Reid~\cite{Reid1}  derived
 a bound on the product of variances of the inferred amplitudes, given by
\begin{equation}
(\Delta_{inf} \widehat{X}_{\theta_1})^{2}(\Delta_{inf}\widehat{X}_{\theta_2})^{2} \geq 1/4
\label{reidineq}
\end{equation}
EPR-steering occurs whenever the above inequality is violated by observables acting on some
given state.


As stated earlier, it is not possible to reveal steering by several continuous variable entangled states
using the Reid inequality (\ref{reidineq}) based on the HUR, even though such states exhibit Bell-violation~\cite{Walborn, PC-TP}. A tighter uncertainty bound is provided by the 
entropic uncertainty relation~\cite{entropy} given by
\begin{equation}
h_{Q}(X) + h_{Q}(P) \geq \ln\pi e
\label{eq8}
\end{equation}
EPR-steering is demonstrated by the non-existence of a LHS model for  measurement outcomes. 
In other words,  EPR steering occurs if the joint measurement probability cannot be written 
as~\cite{Wiseman}
\begin{equation}
\textit{P}(r_{A},r_{B})=\sum_{\lambda}\textit{P}(\lambda)\textit{P}(r_{A}|\lambda)\textit{P}_{Q}(r_{B}|\lambda).
\end{equation}
where $r_{A}$ and $r_{B}$ are the outcomes of measurements $R_{A}$ and $R_{B}$, respectively,
$\lambda$ is the hidden variable,
that specifies an ensemble of states, $\textit{P}$ are general probability distributions and $\textit{P}_{Q}$ are probability distributions which correspond to measurement on the quantum state specified by $\lambda$. The conditional probability $\textit{P}(r_{B}|r_{A})$ is given by
$\textit{P}(r_{B}|r_{A})=\sum_{\lambda}\textit{P}(r_{B},\lambda|r_{A})$ (equivalent to above equation) with $\textit{P}(r_{B},\lambda|r_{A})=\textit{P}(\lambda|r_{A})\textit{P}_{Q}(r_{B}|\lambda) $. Thus,
using the EUR (\ref{eq8}), Walborn {\it et al}.~\cite{Walborn} derived a correspondingly tighter steering condition
given by
\begin{equation}
h(R_{B}|R_{A})+h(S_{B}|S_{A}) \geqslant \ln \pi e 
\label{eq18}
\end{equation}
The violation of the inequality demonstrates steering, as has been explicitly shown
for several Gaussian and non-Gaussian entangled states~\cite{Walborn,PC-TP}.

We now derive a steering criterion based on the uncertainty bound of the sum of variances of
two observables. Let us first consider a typical information theoretic game~\cite{Wiseman}
involving two parties, Alice and Bob.   Alice prepares a bipartite quantum system and sends one 
particle to Bob, and this process can be performed repeatedly. Both the parties can perform
measurements on their respective parts and can communicate classically. Here, Alice's
task is to convince Bob that the state they share is entangled. If, on the other hand, Alice
tries to cheat by sending a pure state drawn at random from an ensemble to Bob, and chooses her
result (communicated to Bob) based on her knowledge this local hidden state (LHS),   the joint probability distribution of their measurement outcomes can be written as~\cite{Wiseman} 
\begin{equation}
P(X_{\theta},Y_{\phi}) = \sum_{\lambda} P(\lambda)P_{Q}(X_{\theta}|\lambda)P(Y_{\phi}|\lambda).
\label{lhs}
\end{equation} 
where $X_{\theta}$ is the observable on Bob's side and  $Y_{\phi}$ is on Alice's side, and
where $P_{Q}(X_{\theta}|\lambda)$ represents the probability of $X_{\theta}$ predicted by a quantum state $\rho_{\lambda}$. 

Our derivation of the steering condition follows the analysis of \cite{Reid1, caval} based on the
HUR.  When Alice  infers the outcomes of Bob's measurement by measuring on her subsystem, the average inference variance of $X_{\theta}$ given the estimate $X_{\theta_{ est}}(Y_{\phi})$ is defined by
\begin{equation}
\Delta_{inf}^{2}X_{\theta} = \langle[X_{\theta} - X_{\theta_{est}}(Y_{\phi})]^{2}\rangle
\end{equation}
where $X_{\theta_{ est}}(Y_{\phi})$ is Alice's estimate of the value of Bob's measurement $X_{\theta}$ as a function of her measurement outcome $Y_{\phi}$, and the average is over all outcomes. The estimate that minimizes the r.h.s. of the above equation is for $X_{\theta_{ est}}(Y_{\phi})=\langle X_{\theta}\rangle_{Y_{\phi}}$~\cite{caval}. 
Thus, the optimal inference variance of $X_{\theta}$ by measurement of $Y_{\phi}$ is given by
\begin{align}
\Delta^{2}_{min}X_{\theta} & =  \langle[X_{\theta} - \langle X_{\theta}\rangle_{Y_{\phi}}]^{2}\rangle \nonumber \\
& = \sum_{X_{\theta},Y_{\phi}} P(X_{\theta},Y_{\phi})(X_{\theta}-\langle X_{\theta}\rangle_{Y_{\phi}})^{2} \nonumber \\
& = \sum_{Y_{\phi}}P(Y_{\phi})\sum_{X_{\theta}}P(X_{\theta}|Y_{\phi})(X_{\theta}-\langle X_{\theta}\rangle_{Y_{\phi}})^{2} \nonumber \\
& = \sum_{Y_{\phi}}P(Y_{\phi})\Delta^{2}(X_{\theta}|Y_{\phi}),
\end{align}
where $\Delta^{2}(X_{\theta}|Y_{\phi})$ is the variance of $X_{\theta}$ calculated from the conditional probability distribution $P(X_{\theta}|Y_{\phi})$, and by definition
\begin{equation}
\Delta_{inf}^{2}X_{\theta} \geq \Delta_{min}^{2}X_{\theta}
\label{inf}
\end{equation}
Assuming the LHS model given by equation(\ref{lhs}), the conditional probability of $X_{\theta}$ given $Y_{\phi}$ can be written as
\begin{align}
P(X_{\theta}|Y_{\phi}) &= \sum_{\lambda} \frac{P(\lambda)P(Y_{\phi}|\lambda)}{P(Y_{\phi})}P_{Q}(X_{\theta}|\lambda)  \nonumber \\
&= \sum_{\lambda}P(\lambda|Y_{\phi})P_{Q}(X_{\theta}|\lambda)
\end{align}
 Since $P(x)$ has a convex decomposition [$P(x)=\sum_{y}P(y)P(x|y)$], the variance $\Delta^{2}x$ over the distribution $P(x)$ cannot be smaller than the average of the variances over the component distribution $P(x|y)$, i.e., $\Delta^{2}x\geq \sum_{y}P(y)\Delta^{2}(x|y)$~\cite{caval}. Then, from the above equation, the variance $\Delta^{2}(X_{\theta}|Y_{\phi})$ satisfies
\begin{equation}
\Delta^{2}(X_{\theta}|Y_{\phi}) \geq \sum_{\lambda}P(\lambda|Y_{\phi})\Delta^{2}_{Q}(X_{\theta}|\lambda),
\end{equation}
where $\Delta^{2}_{Q}(X_{\theta}|\lambda)$ is the variance of $P_{Q}(X_{\theta}|\lambda)$. From the above result it follows that the bound for $\Delta_{min}^{2}X_{\theta} $ is given by
\begin{align}
\Delta_{min}^{2}X_{\theta} &\geq \sum_{Y_{\phi},\lambda}P(Y_{\phi},\lambda)\Delta^{2}_{Q}(X_{\theta}|\lambda) \nonumber \\ &= \sum_{\lambda}P(\lambda)\Delta^{2}_{Q}(X_{\theta}|\lambda).
\label{min}
\end{align}
Hence, for two variables on Bob's side, say $X_{\theta_{1}}$ and $X_{\theta_{2}}$ using Eqs.(\ref{inf}) and (\ref{min})  one has,
\begin{align}
&\Delta_{inf}^{2}X_{\theta_{i}}  \geq \sum_{\lambda}P(\lambda)\Delta^{2}_{Q}(X_{\theta_{i}}|\lambda)
\label{im}
\end{align}
with $i=1,2$. Now, let us define two vectors $u$ and $v$ such that $u \equiv [u_{1},u_{2},..,u_{i},...]$, where $u_{i} = \sqrt{P(\lambda_{i})}\Delta_{Q}(X_{\theta_{1}} |\lambda_{i})$, are the components of the vector $u$, and similarly, $v  \equiv  [v_{1},v_{2},..,v_{i},...]$ with components $v_{i} = \sqrt{P(\lambda_{i})}\Delta_{Q}(X_{\theta_{2}} |\lambda_{i})$.
Noting that $\Delta_{inf}X_{\theta_{1}} \ge |u| (\equiv \sqrt{u_1^2 + u_2^2 + ....})$,
and similarly, for $v$,  in terms of $u$ and $v$, it follows from  Eq.(\ref{im}) that
$\Delta_{inf}X_{\theta_{1}}  \geq |u|$
and 
$\Delta_{inf}X_{\theta_{2}}  \geq |v|$.
Using the triangle inequality ($|u| +|v|\geq|u+v|$) one thus obtains
$\Delta_{inf}X_{\theta_{1}} + \Delta_{inf}X_{\theta_{2}}  \geq \sqrt{(u_1+v_1)^2 + (u_2+v_2)^2 +....}$,
and hence in summation form,
\begin{equation}
\sum_{i=1}^{2}\Delta_{inf}X_{\theta_{i}} \geq 
 \sqrt{\sum_{\lambda}P(\lambda)\left[  \sum_{i=1}^{2}\Delta_{Q}(X_{\theta_{i}} |\lambda)\right] ^{2}}.
 \label{sumsteer1}
\end{equation}

It is known that the quantum fluctuation in the sum of any two observables is always less than or equal to the sum of their individual fluctuations,~\cite{Pati}, i.e.,
\begin{equation}
\Delta(A_{1}+A_{2})\leq \Delta A_{1}+ \Delta A_{2}
\label{un}
\end{equation}
Using the above uncertainty relation (\ref{un}) in the right hand side of Eq.(\ref{sumsteer1}), one gets
\begin{align}
\sum_{i=1}^{2}\Delta_{inf}X_{\theta_{i}} &\geq \sqrt{\sum_{\lambda}P(\lambda)\left[\Delta_{Q}(\sum_{i=1}^{2}X_{\theta_{i}}) |\lambda\right] ^{2}} 
\end{align}

Since we have assumed a LHS model for Bob, the right hand side of the above equation therefore
corresponds to the variance of the sum of the observables $X_{\theta_1}$, and $X_{\theta_2}$, i.e.,
$\Delta(X_{\theta_{1}} + X_{\theta_{2}})$. We thus get the sum steering inequality given by
\begin{equation}
\Delta(X_{\theta_{1}} + X_{\theta_{2}}) \leq \Delta_{inf}X_{\theta_{1}} + \Delta_{inf}X_{\theta_{2}}
\label{eq21}
\end{equation}
A violation of this inequality will demonstrate steering. It may be noted that the variance
of the measured observables on   each individual side will satisfy the uncertainty relation
$\Delta(X_{\theta_{1}}+X_{\theta_{2}})\leq \Delta X_{\theta_{1}} + \Delta X_{\theta_{2}}$.
But, due to the presence of correlations,   Alice's measurement of $Y_{\phi}$ may be
used to infer 
the value of $X_{\theta}$ on Bob's side. Steering takes place if the calculated uncertainties for the inferred observables  violate Eq.(\ref{eq21}). In other words, if the value of $\Delta_{inf}(X_{1}) + \Delta_{inf}(X_{2})$ becomes less than that lower bound of Eq.(\ref{eq21}), we can say that the sum uncertainty relation is able to reveal steering. This is our steering criteria.

\section{Sum steering condition for two qubit Werner states}

Steering for discrete variable systems may be understood by considering an entangled state of two particles, held by two parties (say Alice and Bob)
\begin{equation}
|\Psi\rangle =\sum c_{n}|\psi_{n}\rangle|u_{n}\rangle = \sum d_{n}|\phi_{n}\rangle|v_{n}\rangle,
\end{equation}
where${|u_{n}\rangle}$ and ${|v_{n}\rangle}$ are two orthonormal bases for Alice's system. If Alice chooses to measure in the ${|u_{n}\rangle}$(${|v_{n}\rangle}$) basis, she  instantaneously projects Bob's system onto one of the states$|\psi_{n}\rangle(|\phi_{n}\rangle)$. The
steering analogue~\cite{analog} of the Clauser-Horne-Shimony-Holt (CHSH) inequality  provides a necessary and sufficient criterion for steering in the two measurement per party scenario performed on two-qubit entangled states. The inequality is given as,
\begin{align}
\sqrt{\langle (A_1 + A_2)B_1\rangle^2 + \langle (A_1 + A_2)B_2\rangle^2} \nonumber\\
+ \sqrt{\langle (A_1 - A_2)B_1\rangle^2 + \langle (A_1 - A_2)B_2\rangle^2} \leq 2
\label{achsh}
\end{align}
where, $\lbrace A_1,A_2 \rbrace$ are dichotomic measurements on Alice's side and $\lbrace B_1,B_2 \rbrace$ are dichotomic mutually unbiased qubit measurements on Bob's side. the maximum quantum mechanical value of the left hand side is found to be $2\sqrt{2}$ which can be achieved by Bell states. 

Now, in order to illustrate the EPR steering criterion given by the sum steering inequality (\ref{eq21}) for the case of discrete 
variables, let us consider the  two-qubit Werner state~\cite{Werner} 
\begin{eqnarray}
\rho_W = p |\psi^-\rangle \langle \psi^- | + \frac{1-p}{4} \openone_4
\label{werner}
\end{eqnarray}
where, $|\psi^-\rangle = \frac{1}{\sqrt{2}} (|01\rangle - |10\rangle)$ is the singlet state corresponding to $\sigma_z$-eigenbasis, $\lbrace |0\rangle , |1\rangle \rbrace$, and $\frac{\openone_4}{4}$ is the completely mixed state. The mixing parameter $p$ lies in the range $0\leq p \leq1$. 

Corresponding to two non-commuting spin-$\frac{1}{2}$ observables $\lbrace S_x,S_z\rbrace ,(S_{i}= \frac{\sigma_{i}}{2})$ on Bob's side, the sum steering inequality( \ref{eq21}) becomes,
\begin{eqnarray}
\Delta (S_x^B + S_z^B) \leq \Delta_{\inf} S_x^B + \Delta_{\inf} S_z^B \label{sum}
\end{eqnarray}
The inferred values for the observables $\lbrace S_x,S_z\rbrace$ can be calculated using
the Reid prescription~\cite{Reid1} using the correlation function.  For two general observables
on each side with the correlation function defined as 
$C_{i,j}= \frac{\langle A_iB_j\rangle}{\sqrt{\langle A_i^2\rangle\langle B_j^2\rangle}}$,  the estimates of the variables
$B_1$ and $B_2$ on Bob's side are given in terms of measurements of Alice's variables $A_1$ and $A_2$,
by $B_1^e=g_1A_1, \>\> B_2^e=g_2A_2$, where $g_1$ and $g_2$ correspond to the errors in estimation. The average errors of inference are given by $(\Delta_{inf}A_1)^2= \langle(B_1-B_1^e)^2\rangle =
\langle(B_1-g_1A_1)^2\rangle$, and similarly for $(\Delta_{inf}A_2)^2$. Extremization of the 
inferred errors leads to the conditions $g_1= \frac{\langle A_1 B_1\rangle}{\langle A_1^2\rangle}$
and $g_2= \frac{\langle A_2 B_2\rangle}{\langle A_2^2\rangle}$, which are plugged back into the
expressions for the inferred observables to yield the inferred variances $(\Delta_{inf} B_1)^2$
and $(\Delta_{inf} B_2)^2$. 

For the case of the particular observables chosen here, 
it can be found that, $\Delta_{\inf} S_x^B=\Delta_{\inf} S_z^B=\frac{\sqrt{1-p^2}}{2}$.
 Further, it is found that $\Delta (S_x^B + S_z^B)=\frac{1}{\sqrt{2}}$. Using these results, the sum steering inequality (\ref{sum}) is violated for  $p > \frac{1}{\sqrt{2}}$ which is optimal 
 for any EPR-steering inequality in the two-measurement setting. Fig.~\ref{fig6} depicts the fact that for this range of $p$, the right hand side of Eq.(\ref{sum}) becomes less than the  lower bound
 corresponding to the left hand side. For comparison, we also plot the entropic steering inequality~\cite{entropy1,entropdiscreet}, the left hand side of which for the
 present case becomes
\begin{equation}
H(\sigma_x^B|\sigma_x^A) + H(\sigma_z^B|\sigma_z^A) = -\sum_{+,-}(1 \pm p)\log\frac{1\pm p}{2}
\label{67}
\end{equation}
The above inequality is violated for $p > 0.78$, showing that the entropic steering criterion is not optimal here.

\begin{figure}[ht]
\centering
\begin{minipage}[b]{0.47\linewidth}
\includegraphics[height=3.2cm,width=4.1cm]{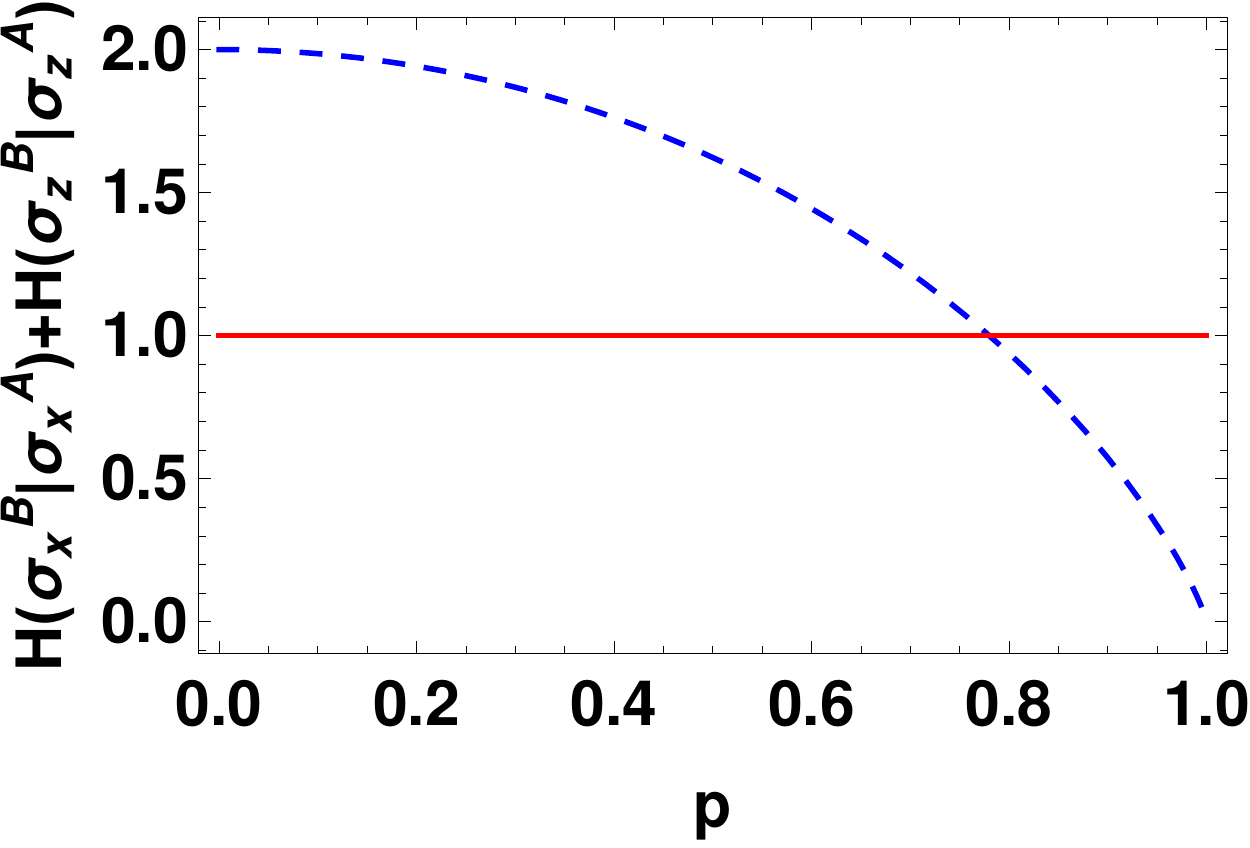}
\end{minipage}
\begin{minipage}[b]{0.47\linewidth}
\includegraphics[height=3.2cm,width=4.1cm]{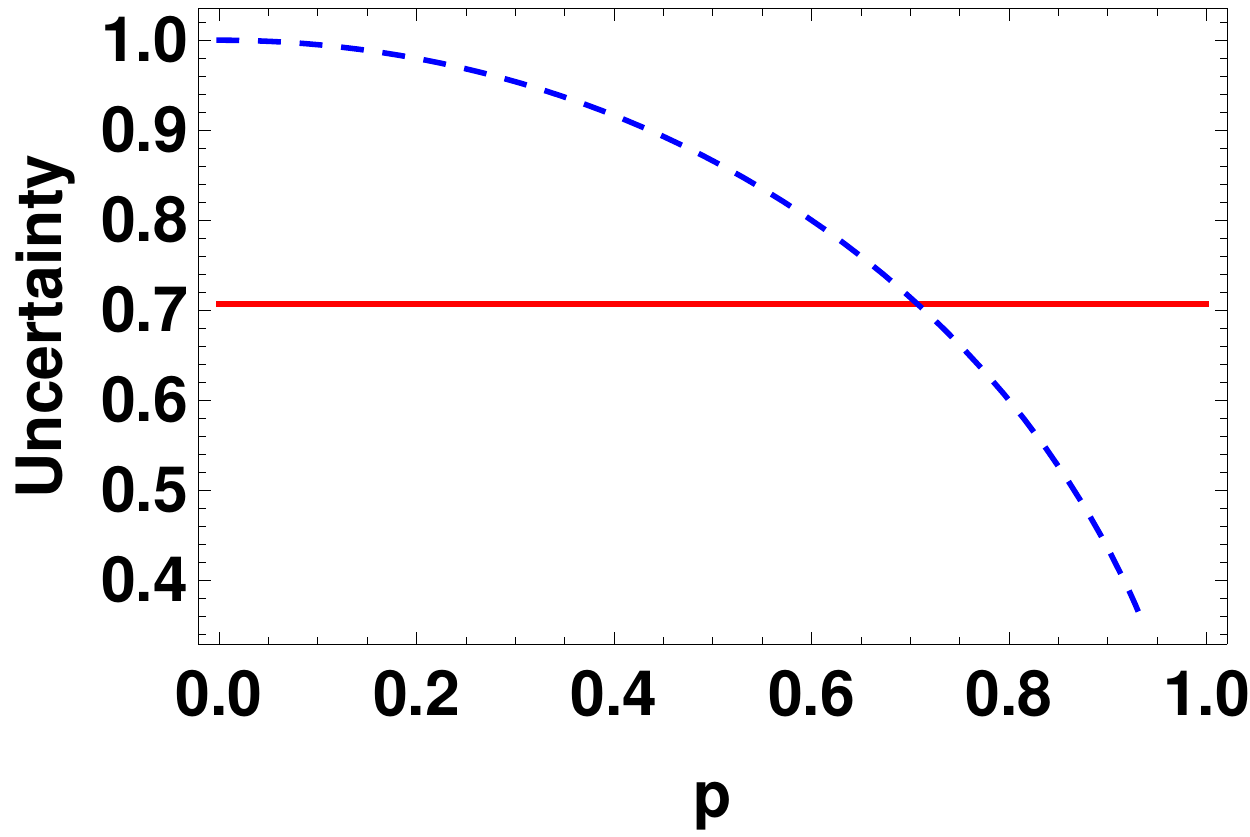}
\end{minipage}
\caption{\textcolor{black}{The left panel displays the entropic steering inequality for spin-$1/2$ observables on the Werner state. The dotted curve is given by Eq.(\ref{67})) while the solid line denotes the lower bound of the entropic steering condition. In the right panel the dotted curve is given by the right hand side of Eq.(\ref{sum}) whereas the solid line represents the lower bound of the sum uncertainty relation which shows steerability for $p > \frac{1}{\sqrt{2}}$.}}
\label{fig6}
\end{figure}

There exist however, other two-measurement steering inequalities~\cite{Saunders}, \cite{caval} that
are violated by Werner states in parameter range, $p > \frac{1}{\sqrt{2}}$. The two-measurement  steering analogue of the CHSH inequality \cite{analog} given by Eq.(\ref{achsh})  provides a necessary and sufficient criterion for steering in the two measurement per party scenario performed on two-qubit entangled states. This steering inequality demonstrates steering iff $p> \frac{1}{\sqrt{2}}$ \cite{arup}. Thus, the optimal range of steerability captured by the inequality given by Eq.(\ref{achsh}) in the discrete variable two-measurement setting is similar to that obtained using
the sum steering inequality (\ref{eq21}) derived  by us.

\section{EPR-steering for continuous variable systems\label{secresult}}

In this section we study steering for various continuous variable states using our sum steering relation and compare our steering criteria with the Reid and entropic steering inequalities.
We consider first the two mode squeezed vacuum state, and then two examples of non-Gaussian entangled states, {\it viz.} the photon subtracted squeezed vacuum state, and the two-dimensional 
harmonic oscillator given in terms of Laguerre-Gaussian (LG) wave functions.
We calculate the magnitude of violation  of our steering inequality and compare it with that 
obtained from the Reid and entropic steering inequalities.

\subsection{Two mode squeezed vaccum\label{Two mode squeezed vaccum}}

The two mode squeezed vacuum state can be generated by applying the two mode squeezing operator $S(\xi) = e^{\xi \widehat{a}_{1}^{\dagger}\widehat{a}_{2}^{\dagger}-\xi^{*}\widehat{a}_{1}\widehat{a}_{2}}$, (where $\xi= re^{i\phi}$) on the two mode vacuum state $|0,0\rangle$, and is given by
\begin{equation}
|NOPA\rangle = |\xi\rangle = S(\xi)|0,0\rangle=\sqrt{1-\lambda^{2}}\sum_{n=0}^{\infty}\lambda^{n}|n,n\rangle
\end{equation}
where $\lambda = \tanh(r)\in [0,1]$, and the squeezing parameter $r> 0$.
The Wigner function corresponding to above state is given by~\cite{Agarwal,PC-TP}
\begin{align}
W_{\xi}(X,P_{X},Y,P_{Y}) \nonumber= \dfrac{1}{\pi^{2}}\exp[&-2(P_{X}P_{Y}-XY)\sinh 2r \\ - (X^{2}+ Y^{2} + P_{X}^{2}& + P_{Y}^{2}) \cosh 2r]
\label{eq23}
\end{align} 
The inferred uncertainty is calculated to be \cite{Reid1,PC-TP},
\begin{equation}
(\Delta_{inf}X_{\theta})^{2} = \dfrac{1}{2} \cosh[2r] - \dfrac{1}{2} \tanh[2r]\sinh[2r]\cos^{2}[\theta+\phi] 
\end{equation} 
We calculate $(\Delta_{inf}X_{\theta_{1}})^{2}$ and $(\Delta_{inf}X_{\theta_{2}})^{2}$ with
the settings $\theta_{1} = 0$, $\theta_{2} = \pi /2$,  $\phi_{1} = 0$ and $\phi_{2} = \pi /2$  (the correlations $\langle XY\rangle$ and $\langle P_{X}P_{Y}\rangle$ are maximized),  
and hence one obtains
\begin{equation}
(\Delta_{inf}X_{\theta_{1}})^{2} = (\Delta_{inf}X_{\theta_{2}})^{2} = \dfrac{1}{2\cosh[2r]}
\end{equation}
Thus, the product of uncertainties $\frac{1}{4\cosh^{2}[2r]}$, is always less than the uncertainty bound ($\dfrac{1}{4}$)~\cite{Reid1} (for $r > 0$). The Reid criterion (\ref{reidineq}) is able to show steering for such states for all squeezing parameters.
One can also apply the entropic steering inequality (\ref{eq18}) for this  state. Since the non-vanishing correlations are $\langle XY\rangle$ and $\langle P_{X}P_{Y}\rangle$, the inequality becomes, \cite{PC-TP}
\begin{equation}
h(\mathcal{X}|\mathcal{Y}) + h(\mathcal{P_{X}|P_{Y}}) \geq \ln \pi e ,
\label{eqNOPA}
\end{equation}
The conditional entropies  are given by
$h(\mathcal{X}|\mathcal{Y}) = h(\mathcal{X},\mathcal{Y}) - h(\mathcal{Y})$
and $h(\mathcal{P_{X}}|\mathcal{P_{Y}}) = h(\mathcal{P_{X}},\mathcal{P_{Y}}) - h(\mathcal{P_{Y}})$ 
with 
$h(\mathcal{X},\mathcal{Y}) = -\int^{\infty}_{-\infty}P(X,Y)\ln P(X,Y)dXdY$,
and similarly for the other entropies. The probability distributions are obtained from the Wigner function (\ref{eq23}). It is already known that entropic uncertainty relation is also able to show steering for all $r$~\cite{PC-TP}.

\begin{figure}[ht]
\centering
\begin{minipage}[b]{0.47\linewidth}
\includegraphics[height=3.2cm,width=4.1 cm]{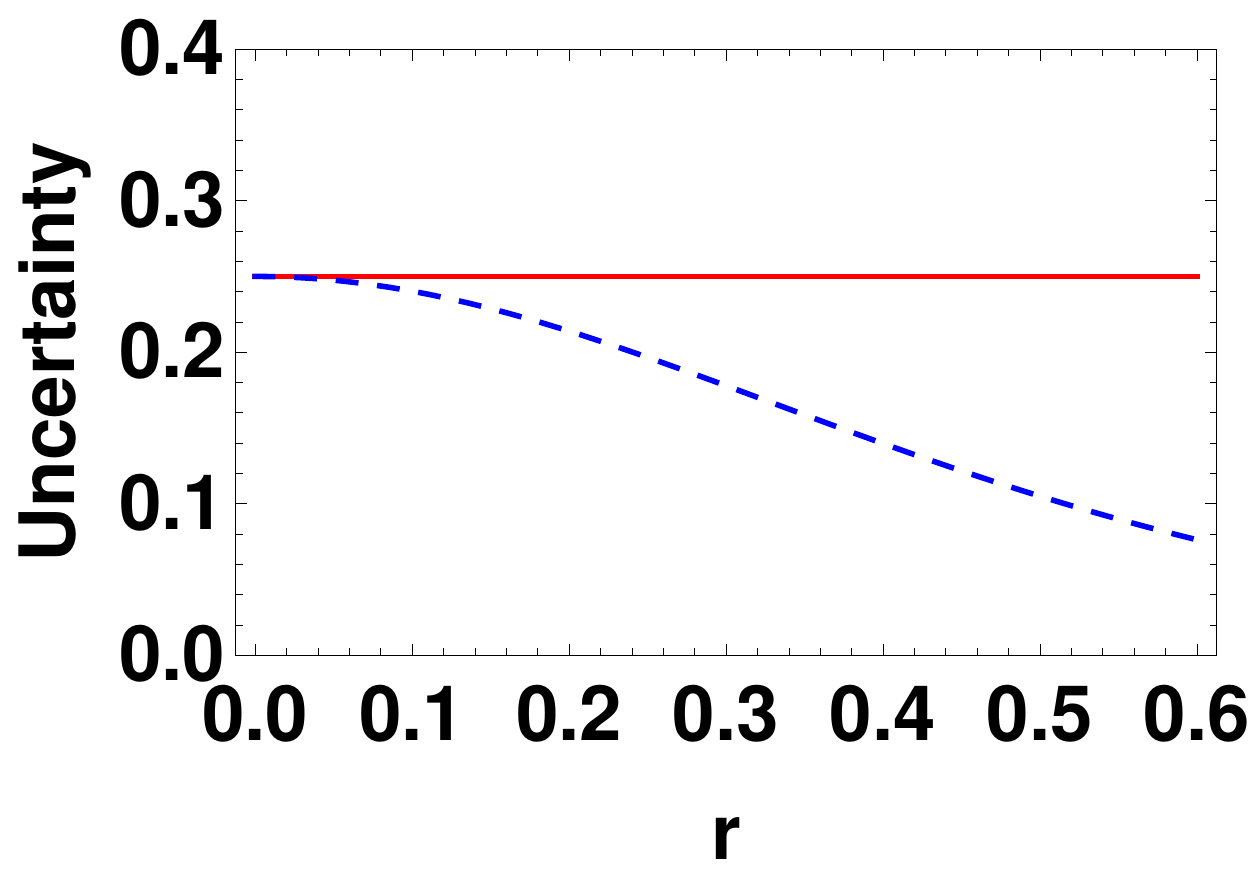}
\end{minipage}
\begin{minipage}[b]{0.47\linewidth}
\includegraphics[height=3.2cm,width=4.1 cm]{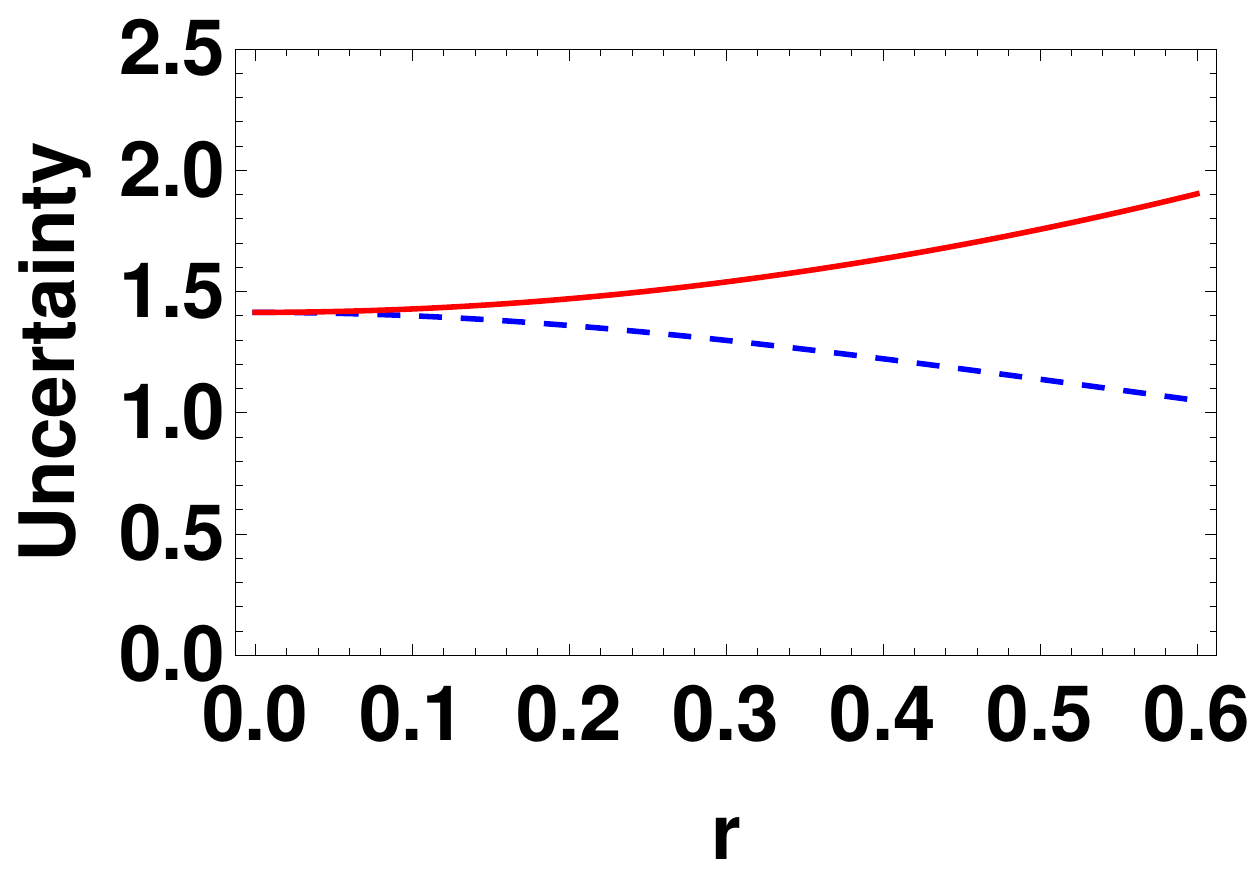}
\end{minipage}
\caption{\textcolor{black}{The left panel shows that the product of inferred uncertainties ($(\Delta_{inf}X_{\theta_{1}})^{2}(\Delta_{inf}X_{\theta_{2}})^{2}$) for the two mode squeezed vacuum state given by the dotted curve falls below the bound (solid line) obtained from the Reid inequality. The right panel shows that the sum of inferred uncertainties (dotted curve) falls below the bound (\ref{sumnopa}) (solid curve) obtained using our sum steering relation.}}
\label{fig1}
\end{figure}

Next, in order to apply our sum steering criterion we calculate $\Delta_{inf}(X_{\theta_{1}}) + \Delta_{inf}(X_{\theta_{2}})$ and the uncertainty bound  we get the value $\Delta(X_{\theta_{1}} + X_{\theta_{2}})$, and get
\begin{eqnarray}
\sum_{i=1}^2 \Delta_{inf}(X_{\theta_{i}})  &= \sqrt{\dfrac{2}{\cosh[2r]}} \nonumber\\
 \Delta( \sum_{i=1}^2 X_{\theta_{i}}) &= \sqrt{2\cosh[2r]}
\label{sumnopa} 
\end{eqnarray}
We plot the uncertainty bound and the inferred uncertainty  in fig(\ref{fig1}). It is clear from the 
figure that our steerability criterion is able to show steering for the two mode squeezed vacuum state.

\subsection{Single photon-subtracted squeezed vaccum\label{Single photon-subtracted squeezed vaccum}}

A non-Gaussian state derived from a two-mode squeezed vacuum by subtracting a single photon from any of the two modes may be written as
\begin{align}
|\xi_{-1}\rangle 
   = \dfrac{1}{2\sinh^{2}(r)}\sqrt{1- \lambda^{2}} & \nonumber \sum_{n=0}^{\infty}\lambda^{n}\sqrt{n}[|n-1,n\rangle 
   +& \\ (-1)^{k}|n,n-1\rangle]
\end{align}
with $k\in [0,1]$.
The Wigner function corresponding to this single-photon subtracted state in terms of $X, P_{X}, Y, P_{Y}$ can be calculated from the Wigner function of the two-mode squeezed vacuum state~\cite{Agarwal}, given by
\begin{align}
\nonumber W_{1}(X,Y,P_{X},&P_{Y})
 = \dfrac{1}{\pi^{2}}\exp[2\sinh(2r)(XY - P_{X}P_{Y})\\
&\nonumber - \cosh(2r)(X^{2}+Y^{2}+P_{X}^{2}+P_{Y}^{2})] \\
&\nonumber\times [ - \sinh(2r)[(P_{X}-P_{Y})^{2}-(X-Y)^{2}] \\
 &+  \cosh(2r)[(P_{X} - P_{Y})^{2}+ (X-Y)^{2}]-1]
\label{eq28}
\end{align}
The uncertainties for the inferred observables $X_{\theta_{1}}$ and $X_{\theta_{2}}$ (for conjugate variables we take $\theta_{1}=0$ and $\theta_{2}=\pi/2$) are given by
\begin{equation}
(\Delta_{inf}X_{\theta_{i}})^{2} = \dfrac{3}{4 [ \cosh(2r) \mp \cosh(r)\sinh(r)] }
\label{suminfphotsub}
\end{equation}

\begin{figure}[ht]
\centering
\begin{minipage}[b]{0.47\linewidth}
\includegraphics[height=32mm, width=41mm ]{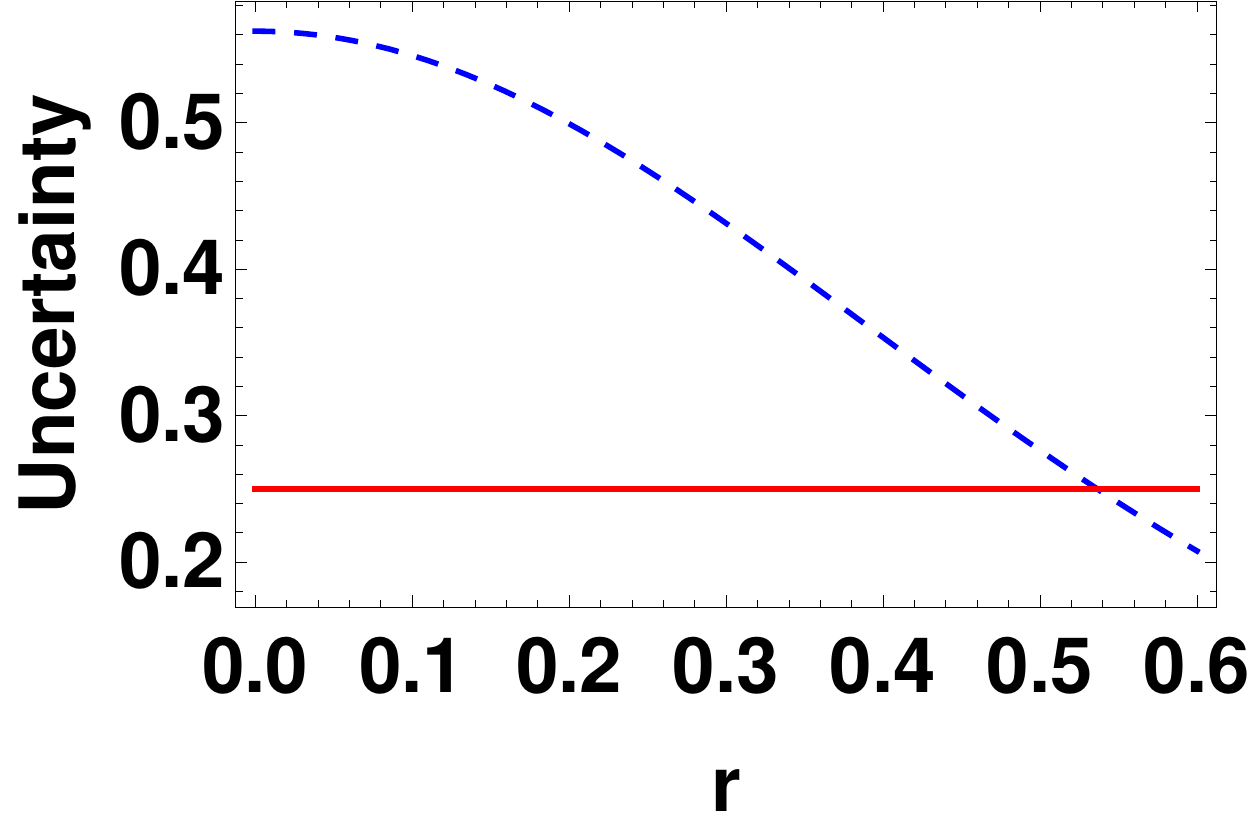}
\end{minipage}
\begin{minipage}[b]{0.47\linewidth}
\includegraphics[height=3.2cm,width=4.1 cm]{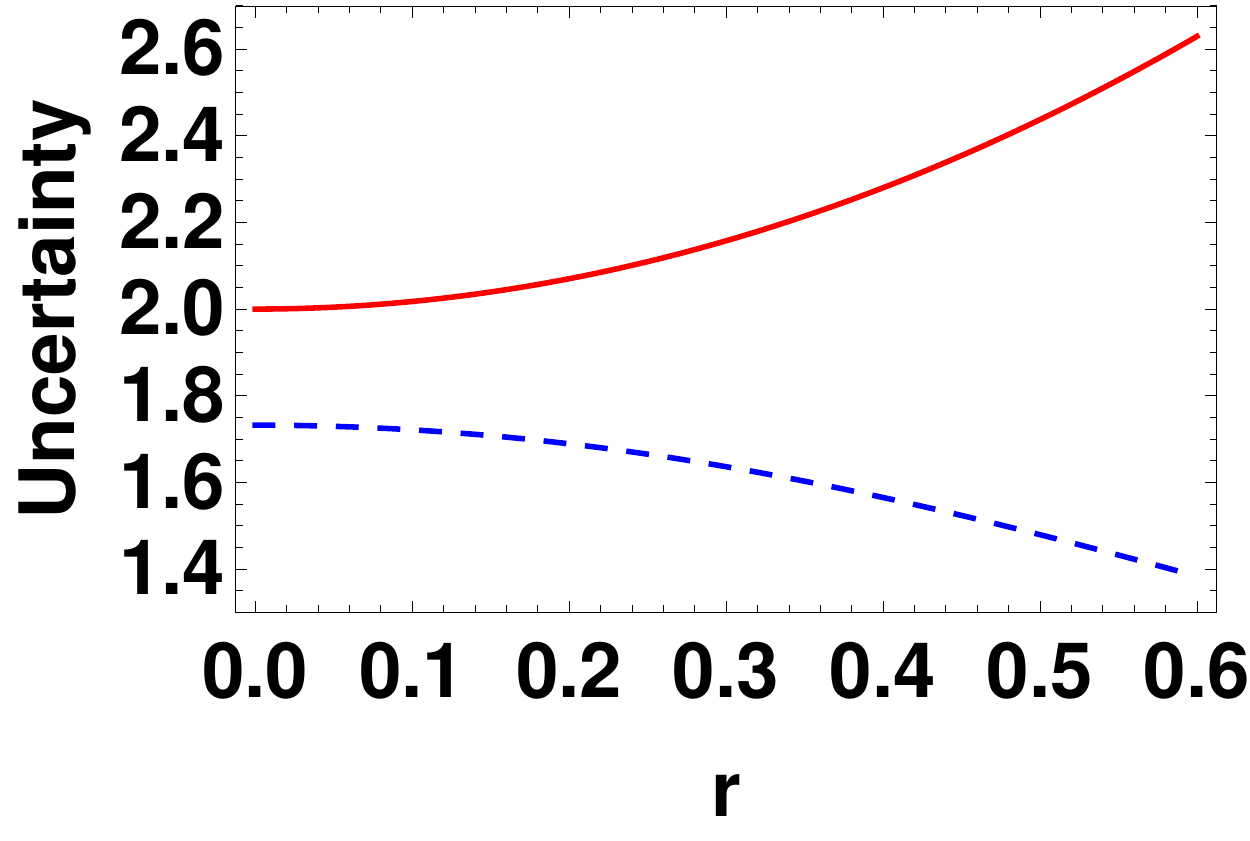}
\end{minipage}
\caption{\textcolor{black}{Steering by the single photon subtracted squeezed vacuum state. On the left hand panel the product of inferred uncertainties ($(\Delta_{inf}X_{\theta_{1}})^{2}(\Delta_{inf}X_{\theta_{2}})^{2}$) (the dotted curve), and the Reid bound (solid line) are plotted versus the squeezing parameter $r$. The right hand panel depicts the sum of inferred uncertainties $(\Delta_{inf}X_{\theta_{1}} + \Delta_{inf}X_{\theta_{2}})$ (the dotted curve) and sum steering bound ($\Delta(X_{\theta_{1}}+X_{\theta_{2}})$)(solid curve). It is clear that the steering inequality based on sum uncertainty relation is able to show steering for all values of the squeezing parameter.}}
\label{fig3}
\end{figure}

(with the minus (plus) sign holding for $i=1(2)$ respectively),
leading to the product of the uncertainties
\begin{equation}
(\Delta_{inf}X_{\theta_{1}})^{2}(\Delta_{inf}X_{\theta_{2}})^{2} = \dfrac{9}{2[3 \cosh(4r)+ 5]}
\end{equation}
A plot of the product of inferred uncertainties and the uncertainty bound given by the Reid criteria with respect to the squeezing parameter $r$ is provided in fig(\ref{fig3}). It is clear from the graph that the Reid inequality fails to exhibit steering for smaller value of $r$, as already known
in the literature~\cite{PC-TP}. The entropic steering inequality (\ref{eqNOPA}) is though able
to reveal steering by this state, as shown earlier~\cite{PC-TP}.

Now, in order to check steering using the sum uncertainty relation,  we  calculate the uncertainty bound for the photon-subtracted state according to sum-uncertainty relation, given by 
\begin{align}
\Delta(X_{\theta_{1}}+X_{\theta_{2}}) = \sqrt{\cosh[2r]-\cos[r]\sinh[r]} \nonumber & \\ +  \sqrt{\cosh[2r]+ \cos[r]\sinh[r]}.
\end{align}
The sum of inferred uncertainties due to the presence of correlations $\langle XY\rangle$ and $\langle P_{X}P_{Y}\rangle$,  is obtained using Eq.(\ref{suminfphotsub}).
We plot the sum uncertainty bound  and the sum of inferred uncertainties for the single photon subtracted squeezed vacuum state in Fig(\ref{fig3}). It is clear from the graph that our steering criterion is able to show steering for all values of the squeezing parameter '$r$'.

Next, we provide a comparison of the magnitude of steering by the three different criteria, {\it viz.}, the Reid criterion, the entropic steering relation, and the sum steering relation for the
single photon subtracted squeezed vacuum state. We present the magnitude of violation for all these steering criteria in the table below, as functions of the squeezing parameter $r$. One can see from the table that the magnitude of violation of the sum steering inequality is always greater than both the Reid and the entropic steering inequalities.
\begin{center}
 \begin{tabular}{||c c c c||} 
 \hline
 r & \tiny $\dfrac{1}{4(\Delta_{inf} \widehat{X}_{\theta_{1}})^{2}\Delta_{inf} \widehat{X}_{\theta_{1}})^{2}}  $ & \tiny $\dfrac{\ln\pi e}{h(\mathcal{X}|\mathcal{P_{Y}}) + h(\mathcal{Y}|\mathcal{P_{X}})}$ & \tiny $\frac{\Delta(X_{1}+X_{2})}{\Delta_{inf}(X_{1}) + \Delta_{inf}(X_{2})}$ \\ [0.5ex] 
 \hline\hline
 0 & 0.444 & 1.044 & 1.155 \\
  \hline
 0.1 & 0.458 & 1.053 & 1.161 \\ 
 \hline
 0.2 & 0.501 & 1.061 & 1.225 \\
  \hline
 0.3 & 0.581 & 1.093 & 1.318 \\
 \hline
 0.4 & 0.707 & 1.124 & 1.457 \\
  \hline
 0.5 & 0.909 & 1.192 & 1.648 \\
 \hline
 0.6 & 1.204 & 1.264 & 1.901 \\[1ex] 
 \hline
\end{tabular}
\end{center}

\subsection{Two-dimensional harmonic oscillator states\label{2d harmonic oscillator}}

For the two-dimensional harmonic oscillator the wave-function in terms of the Hermite-Gauss function is given by~\cite{HG}
\begin{align}
 u_{mn}(x,y)& \nonumber = \sqrt{\dfrac{2}{\pi}}\lbrace \dfrac{1}{2^{m+n}w^{2}m!n!}\rbrace ^{\frac{1}{2}} H_{m}(\frac{\sqrt{2}x}{w})H_{n}(\frac{\sqrt{2}y}{w})\\
&\times \exp[-(x^{2}+y^{2})/w^{2}]
\end{align}
It is possible to construct entangled states using superpositions of the above Hermite-Gaussian
wave functions, that can be represented by Laguerre-Gaussian (LG) beams given by~\cite{Agarwal}
\begin{align}
 \Phi_{mn}(\rho,\theta) &\nonumber= e^{-\rho^{2}/w^{2}}e^{i(m-n)\theta}(-1)^{min(m,n)}\left(\dfrac{\rho\sqrt{2}}{w} \right)^{|m-n|}\\
&\sqrt{\dfrac{2}{\pi m! n! w^{2}}}L^{|m-n|}_{min(m,n)}\left(\frac{2\rho^{2}}{w^{2}} \right)[min(m,n)]!  
\end{align} 
written in terms of cylindrical coordinates  using the generalised Laguerre polynomial.
The Wigner function corresponding to the LG-beam in terms of the dimensionless quadratures $X,P_{X},Y,P_{Y}$ is given by~\cite{simon,PC-TP} 
\begin{align}
 W_{m,n}(X,P_{X},Y,P_{Y}) \nonumber =& \dfrac{(-1)^{(m+n)}}{\pi^{2}}L_{m}[4(Q_{0}+Q_{2})]\\
&L_{n}[4(Q_{0}-Q_{2})]\exp(-4Q_{0})
\label{WLG}
\end{align}
where 
$Q_{0}= \dfrac{1}{4}[X^{2}+ Y^{2}+P_{X}^{2}+P_{Y}^{2}]$,
and 
$Q_{2}= \dfrac{XP_{Y}-YP_{X}}{2}$.
It was shown earlier~\cite{PC-TP} that the Reid criteria  fails to demonstrate steering for LG-beams
However, the entropic steering criterion is able to reveal steerabilty of LG-beams for all values of $n \geq 1$.

We next apply our sum steering inequality (\ref{eq21})  for the case of LG-beams. For this we need to compute the uncertainty bound as well as the inferred uncertainty for the LG-beams. The sum uncertainty bound is obtained in terms of the quadratures, i.e., 
\begin{equation}
\Delta(X_{\theta_{1}}+X_{\theta_{2}})=\Delta(X+P_{X}) = \sqrt{\langle(X +P_{X})^{2}\rangle - \langle X+P_{X}\rangle^{2}}
\label{61}
\end{equation}
and, similarly for the inferred variances
\begin{align} 
(\Delta_{inf}X_{\theta_{1}})^{2} &= (\Delta_{inf}X)^{2} = \langle X^{2}\rangle[1- (C_{0,\pi/2}^{max})^{2}] \nonumber \\
&= \langle X^{2}\rangle \left[ 1- \dfrac{\langle XP_{Y}\rangle}{\sqrt{\langle X^{2}\rangle\langle P_{Y}^{2}\rangle}} \right]
\label{62}
\end{align}
 and
 \begin{align}
 (\Delta_{inf}X_{\theta_{2}})^{2} &= (\Delta_{inf}P_{X})^{2} = \langle P_{X}^{2}\rangle[1- (C_{0,\pi/2}^{max})^{2}] \nonumber \\ 
 &= \langle P_{X}^{2}\rangle \left[ 1- \dfrac{\langle P_{X}Y\rangle}{\sqrt{\langle P_{X}^{2}\rangle\langle Y^{2}\rangle}} \right]
 \label{63}
 \end{align}
The computed values of the above variables are plotted in Fig.3 for several values of the beam
angular momentum $n$, (taking $m=0$). It is seen that steering is demonstrated for all $n\geq 1$.
The violation of the sum steering inequality becomes stronger for higher $n$, a feature that is 
absent in regard to the violation of the entropic steering inequality for LG beams~\cite{PC-TP}.

\begin{figure}[htbp]
  \begin{center}
      \includegraphics[height=5cm, width=8.0cm]{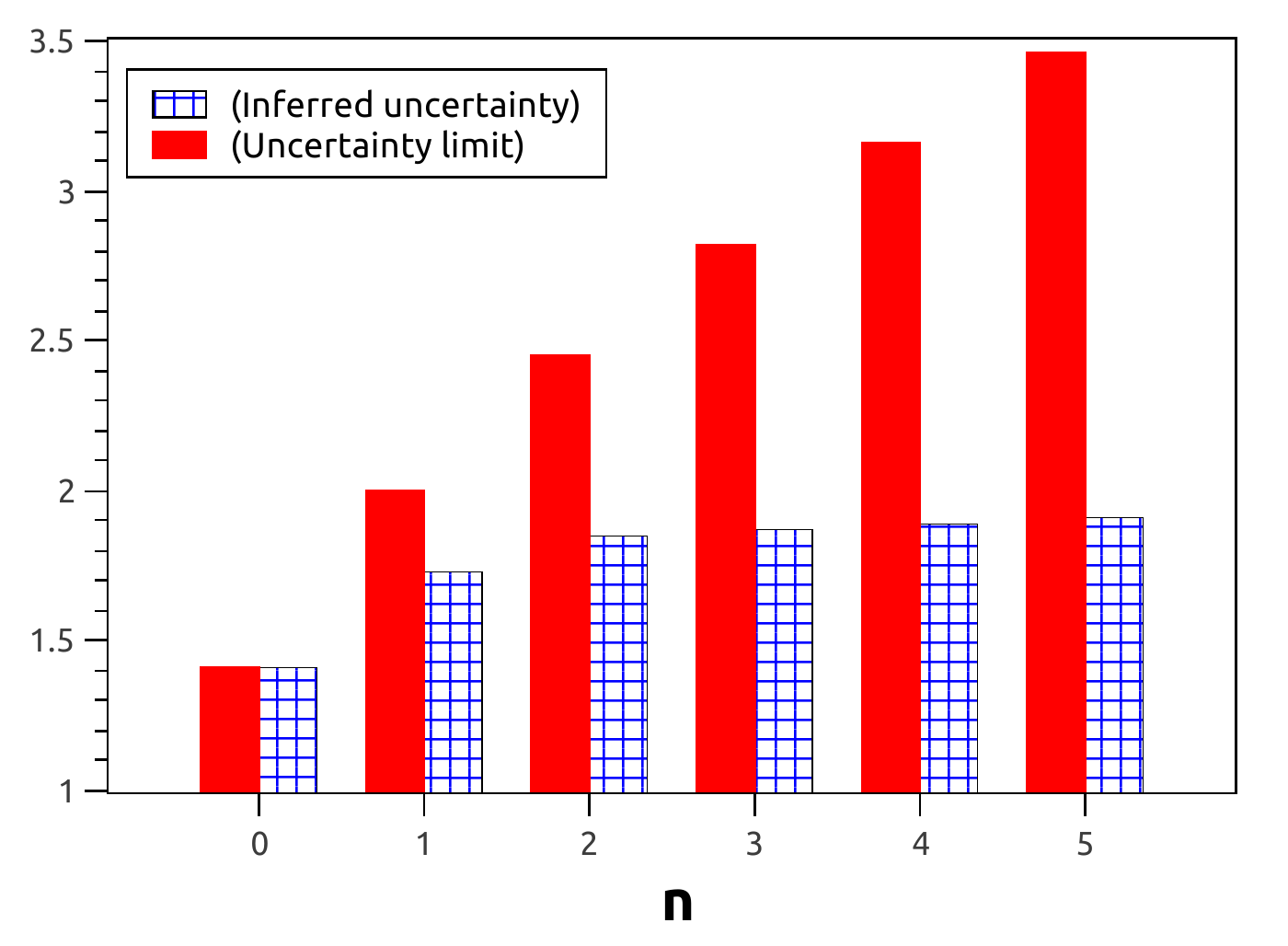}
   \end{center}
\caption{\textcolor{black}{The sum uncertainty bound (solid bar) and the sum of inferred uncertainties (crossed bar) are plotted versus angular momentum $n$ for the LG beams. The figure shows clearly that the violation of the sum steering inequality increases with larger angular momentum.}}
\label{fig6}
\end{figure}

 In the table below we compare the magnitude of violation of our steering inequality with that of the
 Reid inequality and the entropic steering inequality. It is clear from the table that not only does the sum steering relation perform better than the entropic steering inequality for any particular value of $n$ (there is no violation of the Reid inequality), but also that magnitude of violation of the sum steering inequality gets
 stronger with higher angular momentum.
\begin{center}
 \begin{tabular}{||c c c c||} 
 \hline 
 n & \tiny $\dfrac{1}{4(\Delta_{inf} \widehat{X}_{\theta_{1}})^{2}\Delta_{inf} \widehat{X}_{\theta_{1}})^{2}}  $ & \tiny $\dfrac{\ln\pi e}{h(\mathcal{X}|\mathcal{P_{Y}}) + h(\mathcal{Y}|\mathcal{P_{X}})}$ & \tiny $\frac{\Delta(X_{1}+X_{2})}{\Delta_{inf}(X_{1}) + \Delta_{inf}(X_{2})}$ \\ [0.5ex] 
 \hline\hline
 0 & 1 & 1 & 1 \\ 
 \hline
 1 & 0.4444 & 1.0438 & 1.1560 \\
 \hline
 2 & 0.3599 & 1.0567 & 1.3243 \\
 \hline
 3 & 0.3265 & 1.0626 & 1.5080 \\
  \hline
 4 & 0.3086 & 1.0657 & 1.6719 \\
 \hline
 5 & 0.2975 & 1.0676 & 1.8115 \\ [1ex] 
 \hline
\end{tabular}
\end{center}



\section{Conclusions \label{secdiscussion}}

To summarize, in this work we have derived a new steering criterion using the sum uncertainty relation~\cite{Pati,Mac}. Our steering inequality is based on the sum of inferred variances 
pertaining to two
observables of bipartite systems. the sum uncertainty relation leads to a tighter uncertainty
bound compared to the standard product (Heisenberg) uncertainty relation, and hence, the resultant
steering inequality based on the former is expected to yield a tighter steering relation compared
to that obtained from the the latter~\cite{Reid1,Saunders}. In the context of discrete variables, the derived sum uncertainty
based steering relation is able to replicate the steering range of Werner states obtained using the
necessary and sufficient steering condition for the case of bipartite systems with two measurement settings and two outcomes~\cite{analog}. Application of the sum-uncertainty based steering relation for
continuous variable systems demonstrates its advantage over other approaches based on the Reid
criterion~\cite{Reid1} and the entropic steering criterion~\cite{Walborn}. Specifically, we consider examples of non-Gaussian
states such as the photon subtracted squeezed vacuum state and the two-dimensional harmonic oscillator
state to obtain stronger violations of the sum steering inequality compared to those obtained using
the Reid inequality as well as the entropic steering inequality~\cite{PC-TP}. The sum uncertainty based steering
relation hence offers a better prospect of detection of steerability compared to other steering 
criteria for continuous variable systems. It would thus be interesting to explore the practical
feasibility of one-sided device independent key generation~\cite{1sdiqkd} schemes based on the sum steering
relation for continuous variable entangled states.  

\begin{acknowledgments}
We would like to thank Tanumoy Pramanik for his helpful  suggestions. SD acknowledges financial support through INSPIRE fellowship from DST (India). 
\end{acknowledgments}


\end{document}